# CFD Study of Particle Backflow in Pneumatic Conveying Systems due to Triboelectrification.

Otome Obukohwo[a], Andrew Sowinski[a], Poupak Mehrani[a], Holger Grosshans[b*]

[a] Department of Chemical and Biological Engineering, University of Ottawa, 161 Louis-Pasteur, K1N 6N5 Ottawa, Ontario, Canada

[b] Physikalisch-Technische Bundesanstalt. Bundesallee 100, 38116 Braunschweig, Germany

[*] Corresponding author. Email: holger.grosshans@ptb.de, Tel: +495315923510

## Abstract

In industrial plants, pneumatic systems are often used to convey particles from one location to another. Surprisingly, in bench-scale experimental setups, clusters of particles sometimes flow backward or upstream in the conveying channel. In this paper, the effect of electrostatic charge and forces on particle backflow was investigated. Different conveying conditions with varying particle charges were simulated using computational fluid dynamics (CFD), and the resulting flow patterns were compared with CFD simulations of uncharged particles. In a channel flow with periodic boundary conditions in the streamwise and spanwise directions, it was found that electrostatic forces drive particles into low-velocity regions but do not reverse their flow. When transporting particles through a finite-length duct, electrostatic forces cause particles to settle close to the duct's inlet. Finally, when particles were injected into the duct in a pulse, backflow was observed once the particles obtained a charge of 5.04 femto coulombs (fC) or more. The electrostatic forces decelerated the particles at the tail of the pulse and ultimately reversed their direction, whereas the particles at the head of the pulse were accelerated. Thus, it was concluded that electrostatic forces can cause particle backflow in pneumatic conveying systems if particles are fed discontinuously.

## 1. Introduction

Pneumatic conveying is the transport of materials, mostly powders and particulate materials, using a fluid (gas or liquid) as the conveying medium. Literature suggests that materials have been conveyed pneumatically for more than a century. Pneumatic conveying was used to unload grains from ships, as far back as 1856, in ports of London, Rotterdam, Hamburg, and St. Petersburg [1]. Today, it is applied in numerous industries to transport ores, coal, food, powders, and many other materials. For example, it is used in the petrochemical industry to feed catalysts into reactive fluidized beds used in the production of polyethylene.

Pneumatic conveying systems are easier to automate and require less maintenance than mechanical conveyors. On the other side, they need more energy and complex technology to operate [2]. Successful pneumatic conveying requires knowledge of the particles' material characteristics, such as: average size, size distribution, material and bulk densities, moisture content [2], and surface chemistry. All these characteristics, together with the dimensions and surface chemistry of the conveying pipe, affect the conveying velocity of the particles, the pressure drop of the system, and the energy requirements of the system.

The particle velocity is closely tied to the velocity of the conveying liquid or gas. However, particles have occasionally been observed travelling upstream of a conveying line, in the opposite direction of the bulk gas flow [3][4].

Lim et al. [3] simulated a horizontal conveying system, using a computational fluid dynamics-discrete element method (CFD-DEM) approach They simulated (1) a vertical and (2) a 45° inclined pipe with air transporting granules upward against gravity. Backflow was observed when the ratio of electrostatic force to gravitational pull on the particles was greater than 0.5.

In a more recent work, Nimvari et al [4] captured conveyed particles traveling upstream on camera (**Fig. 1**). The pictures in **Fig. 1** show a pipe section close to the outlet of the horizontal conveying system with a transparent pipe that allowed for observation of the powder flow pattern. Backflow of particles was observed when the particles were injected discontinuously in a single pulse. The bulk of the pulse traveled in the direction of the carrier gas, i.e., from left to right. But a fraction of the particles separated from the tail of the pulse. These particles, marked red in **Fig. 1**, formed a cone-like shape and traveled upstream. According to the timestamps and scale, the upstream velocity of the pulse was about 0.04 m/s, a small fraction of the downstream conveying air velocity of 15 m/s.

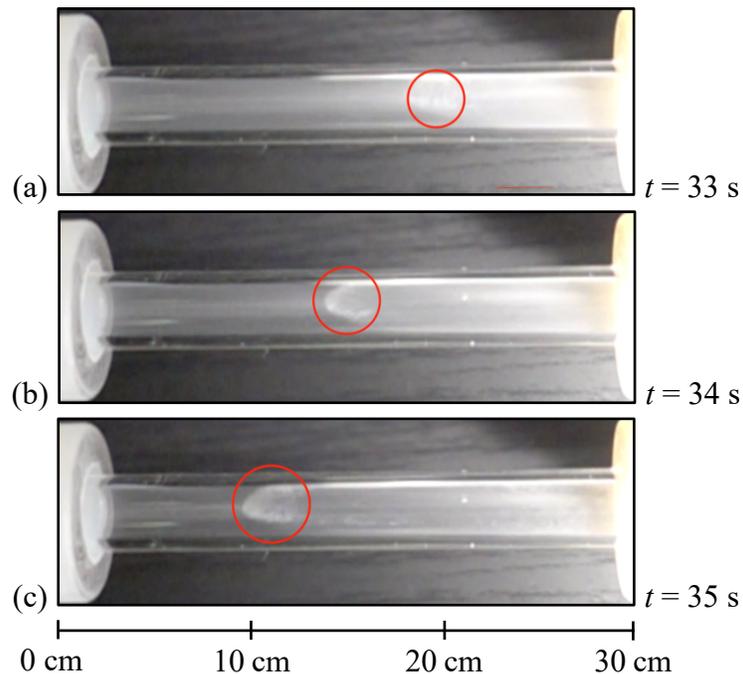

**Fig. 1.** Snapshots of the outlet end of the horizontal pneumatic conveying system at the University of Ottawa. The three pictures were taken one second apart from each other with (a) being the earliest. The carrier fluid flows from left to right, while the marked particle plume moves from right to left, opposing the fluid flow. Unpublished figure from article [4].

The exact reason for particle backflow is still unknown but, electrostatic forces are suspected to play a role [2]. The electrostatic forces are developed by electrostatic charging of particles due to particle-particle and particle-wall collisions, through a process called triboelectrification. These forces are known to cause reduction in the performance of the conveying system by inciting particle agglomeration, adhesion to the conveying walls (known as sheeting, particularly in the polyolefin industry [5]), and increased risk of dust explosions [6].

Literature suggests that the magnitude of electrostatic charge developed increases with an increase in the conveyed particle velocity [6-11], particle diameter [10], pipe length [12], mass loading in continuous particle flow systems, and flow rate of particles [10][13]. On the other hand, increased operational relative humidity [10], pipe diameter [6], and mass loading in pulse conveying systems [5], reduce the magnitude of the developed electrostatic charge.

Measurement of the electrostatic charge carried by individual particles or imposing a defined charge on each single particle are both currently experimentally impossible. Therefore, it is challenging to analyze the effect of electrostatics on particle dynamics in an experiment. Instead, numerical simulations that allow ideal scenarios and initial conditions, like pre-defined particle charges, are attractive methods to investigate the effect of electrostatics, particularly on the observed backflow of particles.

The paper by Lim et al. [3] was focused on the effects of electrostatics on the particle flow pattern. However, the backflow of particles that they observed was assisted by gravity. To our knowledge, there is no other literary work reporting research on particle backflow and electrostatics. Thus, the reason for particles traveling upstream in horizontal conveying lines remains unclear.

This study aims to reproduce the conditions leading to particle backflow and investigate the effect of electrostatics on the backflow. Three types of conveying systems were simulated: (1) a particle-laden channel flow with periodic boundary conditions in the streamwise and spanwise directions; (2) a finite-length duct with continuous particle injection at the inlet; and (3) a finite-length duct with pulsed particle injection. A parametric study for each case started with uncharged particles and proceeded with particles at different charge levels.

## 2. Materials and methodology

### 2.1 Computational Methods

All cases in this paper were simulated with pafiX, a numerical solver for electrostatically charged particle-laden flows. This section summarizes the mathematical models and numerical methods implemented in pafiX. A more detailed explanation has been provided by Grosshans et al. [16].

The solver consists of three parts coupled together: (1) the Navier-Stokes equation to evaluate the fluid flow, (2) Newton's second law of motion to evaluate the movement of the particles, and (3) Gauss's law to evaluate the electrostatic field. These three parts are summarized below.

The fluid motion of the flow is described by the Navier-Stokes equations for incompressible flows,

$$\nabla \cdot \boldsymbol{u} = 0 \tag{1}$$

$$\frac{\partial \boldsymbol{u}}{\partial t} + (\boldsymbol{u} \cdot \nabla)\boldsymbol{u} = -\frac{1}{\rho}\nabla p + \nu \nabla^2 \boldsymbol{u} + F_\text{s} + F_\text{f}, \tag{2}$$

where $u$ is the fluid velocity, $\rho$ is the fluid density, $p$ is the fluid dynamic pressure, and $\nu$ is the fluid kinematic viscosity. $F_s$ is the source term representing the momentum transfer from the particles to the fluid, and $F_f$ balances momentum loss due to friction at the walls.

Each particle position is tracked in a Lagrangian framework by solving Newton's second law of motion,

$$\frac{du_p}{dt} = f_{fl} + f_{coll} + f_{el}, \qquad (3)$$

where $u_p$ is the particle velocity, $f_{fl}$ is the specific aerodynamic force exerted by the surrounding fluid, $f_{coll}$ is the specific force on the particle because of particle-particle or particle-wall collisions, and $f_{el}$ is the specific force due to the electrostatic field.

The specific aerodynamic force is calculated as:

$$f_{fl} = -\frac{3\rho}{8\rho_p r} C_d |u_{rel}| u_{rel}, \qquad (4)$$

where $\rho_p$ is the particle density, $C_d$ is the particle drag coefficient and $u_{rel}$ is the particle velocity relative to the fluid.

The specific collision force is calculated only when two particles make contact, or when a particle reaches a wall. Details on the algorithm for calculating $f_{coll}$ can be found in the report by Grosshans et al. [16].

Finally, the specific electrostatic force is calculated as:

$$f_{el} = \frac{QE}{m_p}. \qquad (5)$$

In **Eqn. 5**, $Q$ represents the particle's charge, $m_p$ its mass, and $E$ is the electric field strength around the particle. This electric field strength follows Gauss's law and is defined in terms of the electric potential as:

$$E = -\nabla \varphi_{el}, \qquad (6)$$

where,

$$\nabla^2 \varphi_{el} = -\frac{\rho_{el}}{\varepsilon}. \qquad (7)$$

Herein, $\varphi_{el}$ is the electric potential. $\rho_{el}$ is the electric charge density, and $\varepsilon$ is the electric permittivity. For this system of a low solid volume fraction, $\varepsilon$ is approximated by the free space permittivity (8.85 x $10^{-12}$ Fm$^{-1}$).

## 2.2 Simulation Setup

The effect of electrostatics was investigated by a three-fold study considering three different numerical set-ups. First, flow through a channel was simulated (left on **Fig. 2**). This set-up was chosen to test the general effect of electrostatic forces on the particle velocity, independent of geometrical constraints and inlet or outlet conditions. The channel dimensions were 0.24 m × 0.04 m × 0.08 m, in the $x$, $y$ and $z$ directions respectively; the domain was discretized by a computational grid containing 256 × 144 × 144 cells in the $x$, $y$, and $z$ directions, respectively. The size of the cells in the grid was not constant because a stretch grid was used. The cells were finer close to the walls to provide better resolution. A mesh independence study for a similar boundary size has been reported by Grosshans et al. [16]. The minimum mesh size was found to be 240 × 140 × 140 cells, which is smaller than the mesh used in all simulations presented in this paper.

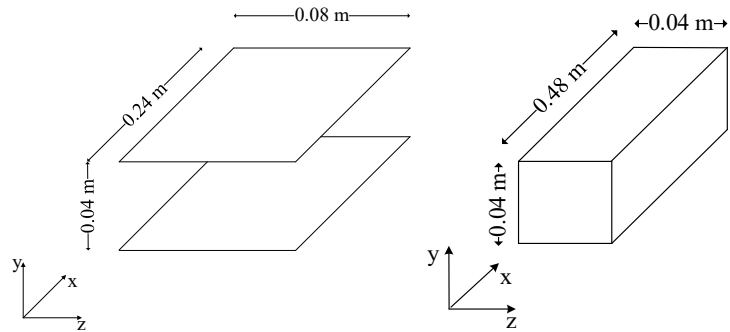

**Fig. 2.** Schematics of the computational domain of the channel (left) and the duct (right).

Second, flow through a finite-length duct (right on **Fig. 2**) with square-shaped cross-section was simulated. Particles were fed into the inlet at a constant feeding rate and left the duct at the outlet. The aim of this step was to find out whether an inlet and outlet boundary condition is related to the occurrence of backflow. Third, flow through a finite-length duct was simulated again but, this time, the particles were injected in pulses. The aim of this step was to check whether discontinuous injection could cause backflow. The duct dimensions were 0.48 m × 0.04 m × 0.04 m, discretized by 512 × 144 × 144 cells, in the $x$, $y$ and $z$ directions. The number of cells in the $x$ direction was doubled, along with the length in the $x$-direction, to keep the cell density constant between the channel and duct simulations.

The focus of the simulations was the interplay of fluid mechanics and electrostatics, so gravitation was neglected. All walls, in both channel and duct simulations, were assumed to be conductive and grounded, consequently, zero-Dirichlet boundary conditions were applied to the electric potential.

Further, all simulations investigated the same fluid and particle properties: a fluid flow of a frictional Reynolds number ($Re_\tau$) of 180 and a particle with a Stoke's ($St$) number of 8. The particles were monodispersed and spherical with a diameter equal to 15 μm and a density equal to 9480 kgm$^{-3}$. These properties were chosen to achieve a Stoke's number equal to 8. The focus of the paper was on the role of electrostatics in the particle backflow phenomenon, and the conditions required to incite backflow. So, the particle properties were reduced to a dimensionless number (Stoke's number) and kept constant.

The $Re_\tau$ and $St$ are representative of realistic pneumatic conveying systems, the Reynolds number being at the lower end and the Stokes number at the upper end of the typical operating range [20]. The fluid kinematic viscosity and density were 1.46 m$^2$s$^{-1}$ and 1.2 kgm$^{-3}$, respectively.

In the channel flows, $10^8$ particles were seeded per unit volume. Because of the periodic boundary conditions, no particle was lost during the simulation. In the finite-length duct, we injected $5 \times 10^5$ particles per second, to obtain a particle density comparable to the density in the channel flows. For the third part of the study with pulse injections, the seeding pulses resembled square waves with two parameters, $t_1$ and $t_2$, defined as the duration of a pulse and the time interval between two consecutive seeding pulses, respectively. The schematic and definition of the square wave are shown in **Fig. 3**, **Eqn. 8** and **Eqn. 9**.

$$\dot{N}_{in}(t) \; (s^{-1}) = \begin{cases} 5 \times 10^5 & t - t_{ref} < t_1 \\ 0 & t - t_{ref} > t_1 \end{cases} \tag{8}$$

$$t_{ref} = \left\lfloor \frac{t}{t_1 + t_2} \right\rfloor (t_1 + t_2) \tag{9}$$

In **Eqn. 9**, the operator $\lfloor x \rfloor$ denotes the floor function that returns the largest integer less than $x$. The range of $t_1$ was determined as a function of the centerline velocity of the fluid at the inlet, $u_{bulk,t}$, and the width of the duct, $y_{dim}$. The range of $t_2$ was determined as a multiple of $t_1$

$$t_1 = \tau_1 \frac{y_{\text{dim}}}{u_{bulk,t}} \qquad (10)$$

$$t_2 = \tau_2 t_1 \qquad (11)$$

Ultimately, the non-dimensional parameters $\tau_1$ and $\tau_2$ were used to define the injection pulses. For each of the three set-ups, we simulated one system with uncharged particles, and multiple systems with different charge levels. The charge levels, and the electric Stoke's number of the particles, ($St_{\text{el}}$), are shown in **Table 1**. $St_{\text{el}}$ values were calculated as the ratio of the particle timescale to the electrostatic timescale.

**Table 1.** List of simulations with charger per particle and electrostatics Stoke's number ($St_{el}$).

| Sim No. | Charge (fC) | $St_{\text{el}}$ ($10^{-3}$) |
|---|---|---|
| 1 | 0.00 | - |
| 2 | 0.63 | 1 |
| 3 | 1.26 | 2 |
| 4 | 2.52 | 4 |
| 5 | 5.04 | 8 |
| 6 | 50.40 | 80 |
| 7 | 252.00 | 400 |
| 8 | 504.00 | 800 |
| 9 | 5040.00 | 8000 |

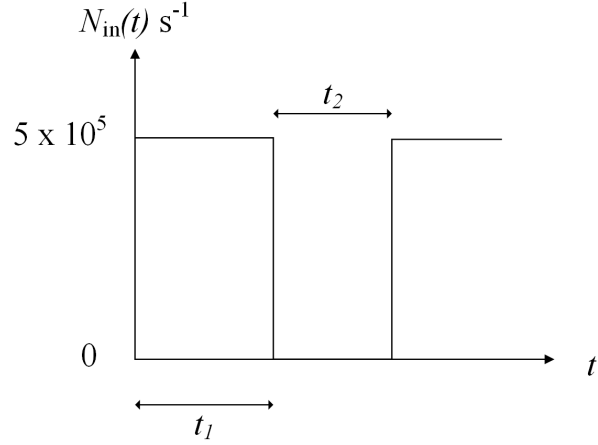

**Fig. 3.** Schematic description of the pulse wave.

## 3. Results and Discussion

### 3.1 Channel flow with periodic boundary conditions

In this section, the results from channel flow simulations are presented. The particles were seeded randomly at the start of each case, and the simulation proceeded until the flow became statistically stationary.

To analyze the effect of particle charge on the particle velocity profile, the simulation boundary was divided into bins along the $y$-direction (wall normal direction). The mean streamwise velocity of particles within each bin was calculated and normalized with the friction velocity, $u_\tau$ to obtain $<u_p^+>$. $u_\tau$ was calculated as:

$$u_\tau = Re_\tau * \frac{\nu}{0.5 * y_{\text{dim}}} \tag{12}$$

with $y_{\text{dim}}$ being equal to the width of the channel.

**Fig 4.** Shows the average streamwise particle velocity as a function of the distance of the particles from the walls. The distance of the particles from the wall is presented in wall units, $y^+$ (dimensionless wall distance parameter) on the horizontal axis. To enlarge the flow region close to the wall, the horizontal axis is in the logarithmic scale. The normalized mean velocities were plotted up until the center of the channel due to the symmetry of the flow patterns.

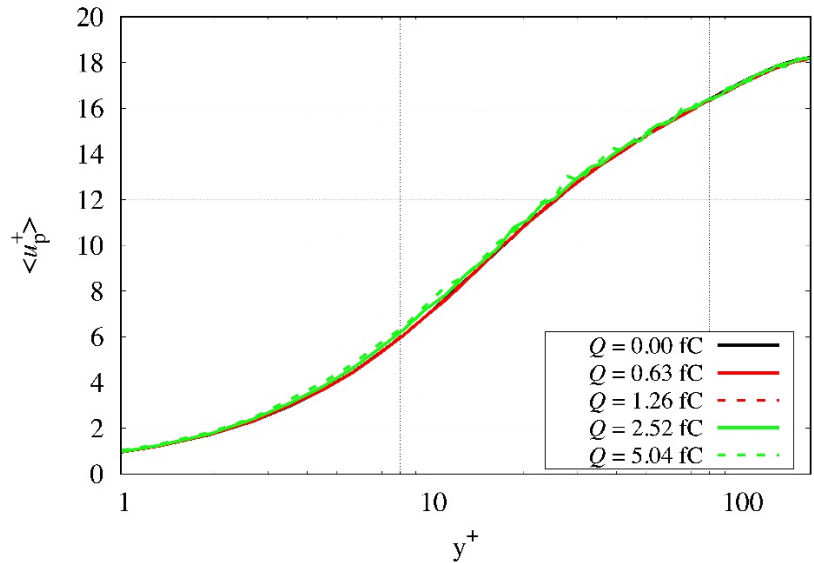

**Fig. 4.** Average streamwise particle velocity profiles depending on the distance from the wall for uncharged and charged channel flow simulations. All cases showed identical velocity averages suggesting no net streamwise effect of electrostatic force on particles.

No significant difference was observed in the velocity profiles between particles holding different charges. This suggests that there was no net effect of the electrostatic force in the streamwise direction. The periodic boundary implies an infinite length duct in front of, and behind, any particle regardless of its position. The temporal average of electrostatic force due to particles ahead and before a particle cancel out. Hence, the sum of the electrostatic forces, in the streamwise direction on the particle is zero.

**Fig. 5** shows the normalized number density of particles, $C$, as a function of the distance from the wall in wall units, $y^+$. Again, the simulation boundary was divided into bins along the $y$-direction (wall normal direction). The number of particles in each bin was calculated and divided by the volume of the bin to produce a bin particle density. The bin particle density was normalized by the overall particle density (number of particles in the entire volume) to produce a normalized density $C$.

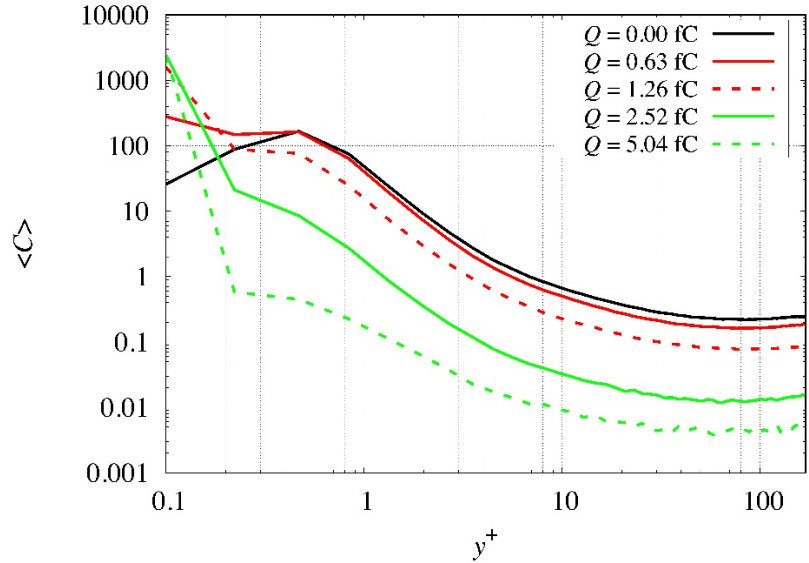

**Fig. 5.** Normalized particle number density (concentration) as a function of the distance (in wall units) from the walls for uncharged, 0.63 fC, 1.26 fC, 2.52 fC and 5.04 fC channel flow simulations. Both horizontal and vertical axes are plotted in the logarithmic scale. Curves show particle migration toward the walls with increase in charge per particle.

**Fig. 5** shows that particles migrated to the walls with as the particle charge increased. The uncharged particles were distributed most uniformly, and the accumulation of particles at the walls increased with increase in the particle charge. This implied a net electrostatic force effect, in the wall-normal direction, on the particles, in this case: an attractive force to the walls. In all cases, but the uncharged simulation, the concentration peaked at the wall. This can be explained by turbophoresis: migration of particles towards the area with the lowest turbulence level. The non-wall peak in the uncharged case has also been observed by Gizem et al. [15].

Even though the particles' velocity profile is independent of their charge as seen in **Fig. 4**, electrostatic charge heavily affected the volume-average streamwise particle velocity, see **Fig. 6**. Uncharged particles moved with a volume-average streamwise velocity of about 4.5 m/s downstream. Adding charge to the particles reduced the velocity to values lower than 0.3 m/s for charge levels of 2.52 fC and higher. The decline observed in **Fig. 6** is harmonious with the increase in concentration of particles closer to the channel walls. The particles moving close to the wall entered the fluid's low velocity region, decelerated, and adopted the fluid's velocity, becoming nearly immobile.

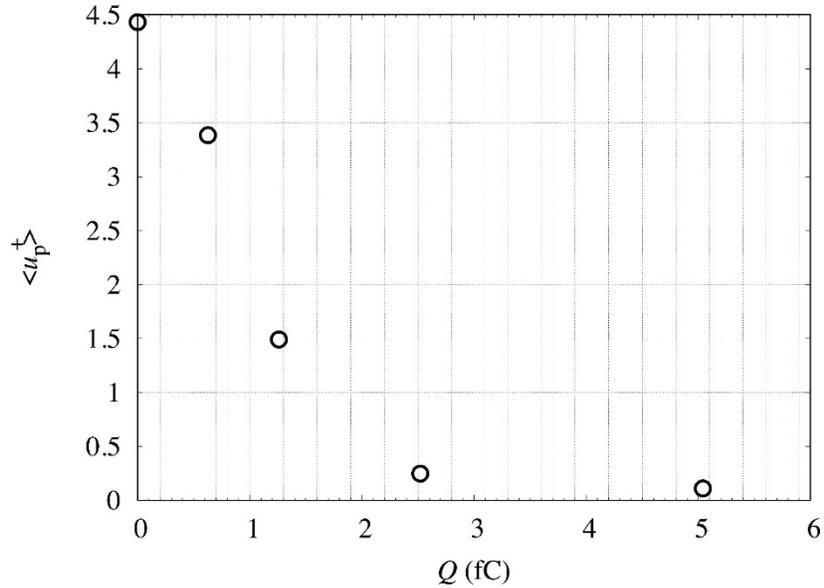

**Fig. 6.** Volumetric average of particle streamwise velocities as a function of charge per particle in channel flow simulations. Decline in mean velocity with increase in charge confirms notion of particle migration toward low velocity regions.

In summary of **Fig. 4, 5 and 6**, the streamwise particle velocity at a specific distance from the wall was independent of particle charge. But, charged particles tended to migrate to low-velocity regions; thus, the volume-average streamwise particle velocity decreased with increase in particle charge. These observations suggest that with greater electrostatic charge, the volume-average velocity will get asymptotically close to zero: all particles will move towards the walls into lower velocity zones and become nearly immobile.

Finally, **Fig. 7** compares the probability distribution of the velocities for all simulations. The comparison shows an increase in the probability of lower particle velocities with an increase in particle charge. In both **Fig. 6** and **Fig. 7**, the effect of particle migration to the wall is seen. However, **Fig. 7** proves that all particles in the system, independent of their charge, travel with a positive downstream velocity and backflow does not occur.

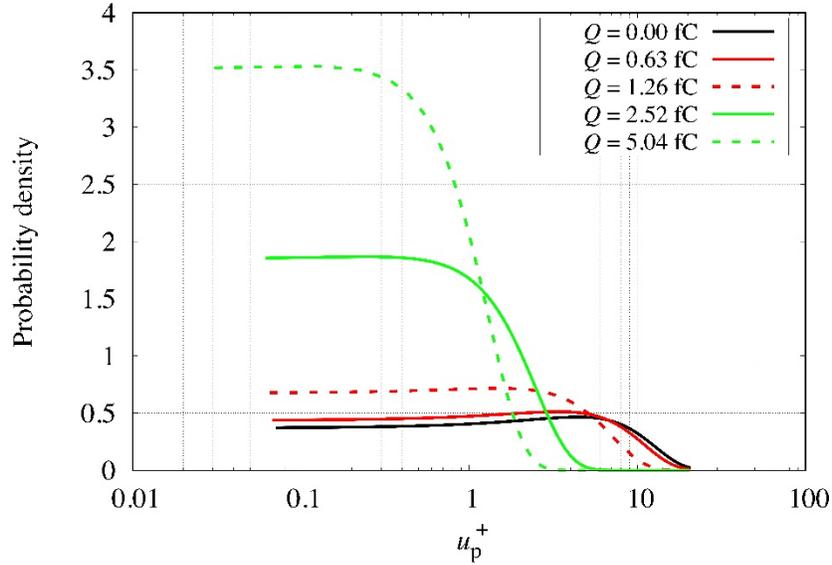

**Fig. 7.** Velocity distribution at statistical convergence for uncharged, 0.63 fC, 1.26 fC, 2.52 fC, and 5.04 fC simulations of channel flow. Increase of probability density of low particle velocities with charge confirms notion of migration of particles to lower velocity zones.

In conclusion, an increase in particle charge causes an increase in the electric field strength and consequently the strength of the attractive force on particles toward the walls. This leads to migration of particles toward the walls and into lower velocity zones. The increased number of slow-moving particles causes a decrease in the volume-average particle velocity in the conveying system. The absence of net streamwise electrostatic force, and consequently no backflow, can be explained by the equality, on average, of the temporal electrostatic force experienced by a particle from the particles behind it and the particles in front of it. This equality will be called streamwise symmetry from this point on.

### 3.2 Duct flow with continuous particle injection

This section presents the results from the finite-length duct flow simulations with continuous seeding. In these simulations, the particles were fed into the system at a rate of 5 x $10^5$ particles per second. The simulations proceeded until the flow became statistically stationary.

Analogous to **Fig. 4**, **Fig. 8** plots the mean streamwise particle velocity, $< u_p^+ >$, as a function of distance from the wall (in the $y$-direction) $y^+$ at each different charge level. As flow in a duct is symmetric across both spanwise and vertical planes [16], **Fig. 8** shows only half of the duct.

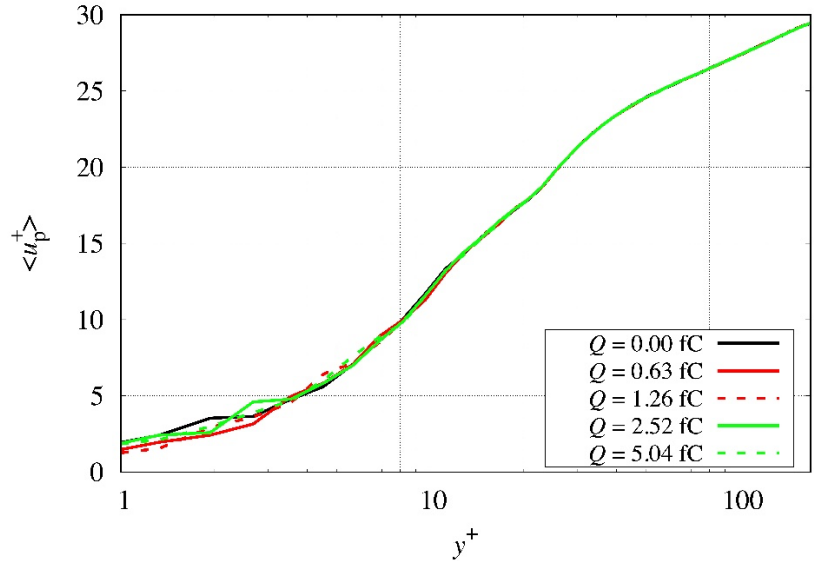

**Fig. 8.** Average streamwise particle velocities at different wall units away from the wall for uncharged, 0.63 fC, 1.26 fC, 2.52 fC and 5.04 fC continuous seeding inlet-outlet duct flow simulations. All cases showed identical velocity averages suggesting no net streamwise effect of electrostatic force on particles.

No significant difference was observed among different charge levels in **Fig. 8** and analysis of the concentration of particles, as a function of distance from the wall, **Fig. 9(a)**, disclosed an increased migration of particles toward the walls as the particle charge increased. As was seen in **3.1**, there was no net effect of electrostatic force in the stream wise direction and, the particles experienced an attraction to the wall in the presence of charge, with the attraction growing with the charge. However, unlike the channel flow, there was no clear indication of this migration in the volume-average streamwise velocity (**Fig. 9(b)**). The volume-average velocity did not change significantly across the different charge levels. The absence of this trend can be explained by the continuous seeding of new particles at the inlet. The newly seeded particles have velocity distributions in-line with the velocity distribution of the fluid and so contain faster moving particles that balance out the particles that migrate into low velocity zones.

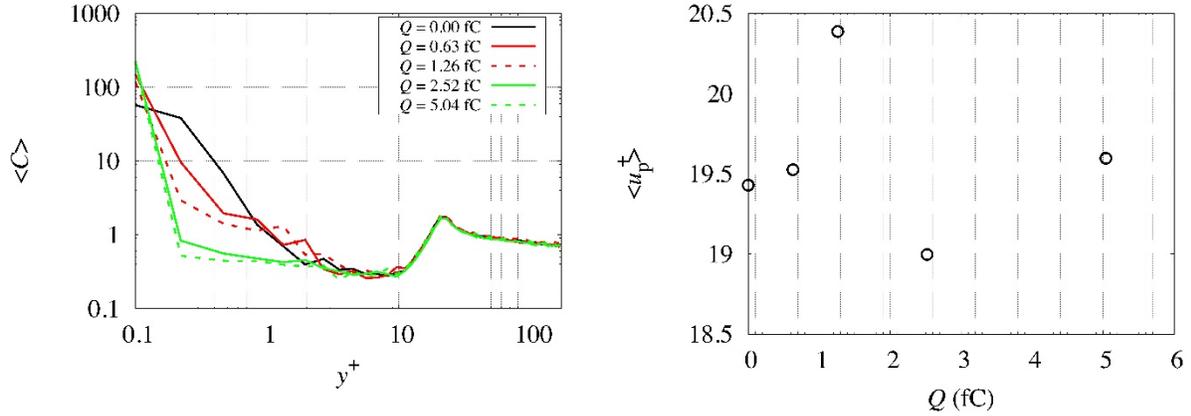

**Fig. 9.** (a) Normalized particle number density as a function of the distance in wall units from the walls for the uncharged, 0.63 fC, 1.26 fC, 2.52 fC and 5.04 fC continuous duct flow simulations showing a migration of particles to the walls with increase in charge per particle. (b) Volumetric mean particle streamwise velocity as a function of the charge per particle in continuous duct flow simulations showing no significant effect of particle migration on mean velocity.

According to **Fig. 9(a)**, higher concentrations appear closer to the wall ($y^+ < 0.2$) in the 2.52 fC simulation than in the 5.04 fC. This can be explained by particles bouncing of the walls due to a larger attractive acceleration in the 5.04 fC case.

Further, the wall-detached peak observed in the uncharged channel flow is absent. This observation can be explained by the difference between the type of convergence reached in **3.1** and **3.**2. The channel flows with periodic boundary conditions were simulated until the flow was fully developed in space and time. On the contrary, the duct flows were simulated until the flow reached steady state in time alone because the length of the duct of 0.48 m is not sufficient for the flow to fully develop in space. Thus, the finite length of the duct does not suffice to form the wall-detached concentration peak.

In conclusion, in the finite-length duct simulations, the particles migrated towards the walls and into low velocity zones as their charge increased. This migration was not reflected in the volume-average streamwise velocity as new faster particles were continuously fed into the duct. There was no backflow of particles up to an $St_{el}$ value of $8 \times 10^{-3}$ (**Fig. 10**). The absence of net streamwise electrostatic force, and consequently no backflow, can again be partly explained by the streamwise symmetry in the simulation.

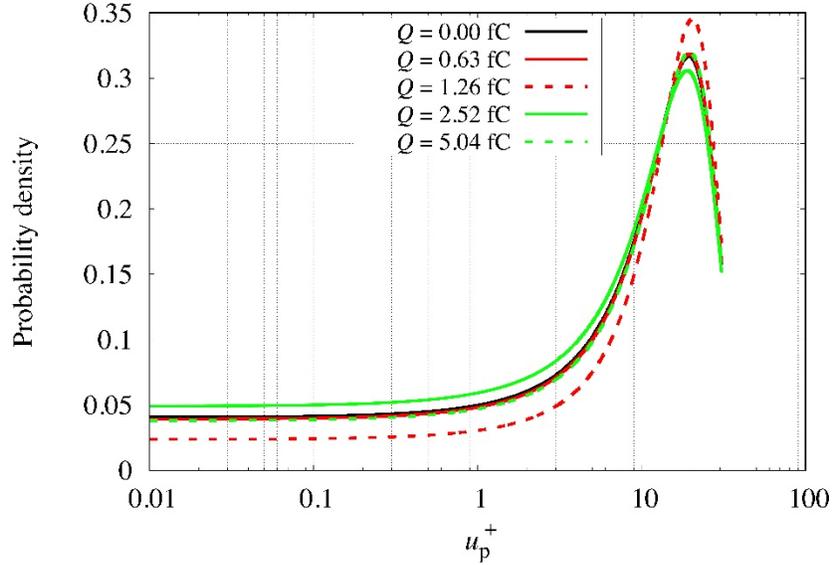

**Fig. 10.** Velocity distribution at statistical convergence for uncharged, 0.63 fC, 1.26 fC, 2.52 fC, and 5.04 fC simulations of continuous seeding duct flow. Increase of probability density of low particle velocities with charge confirms migration of particles to lower velocity zones.

### 3.3 Duct flow with pulsed particle injection

This section presents the results from the finite-length duct flow simulations with pulsed seeding. The simulation setup for this phase is identical to the setup presented in **3.2**. However, as opposed to continuous particle injection, one plume of particles was seeded. Then, the simulations proceeded until the particles left the system.

Contrary to the stream symmetry suggested in **3.1**, particles in a plume are expected to experience different net streamwise electrostatic forces. The particles at the tail of the plume may be pushed back, the particles in the middle of the plume may have no net streamwise electrostatic force, and the particles at the head of the plume may be pushed forward. Hence, the electrostatic force is hypothesized to be negative (relative to the gas flow directions) at the rear end of the plume and positive (relative to the gas flow direction) at the front end of the plume with a positive gradient along the plume from the tail to head. To check this hypothesis, the uncharged simulation was compared with the 5.04 fC simulation in **Fig. 11**.

The charged case contained more slow-moving particles at $t = 0.11$ s and $t = 0.14$ s due to deceleration by electrostatic forces. As a result of repulsion between the like-charged particles, at

$t = 0.14$ s, the plume area at the outlet (right end of pictures) of the charged case had a larger width (in vertical direction) than in the uncharged case.

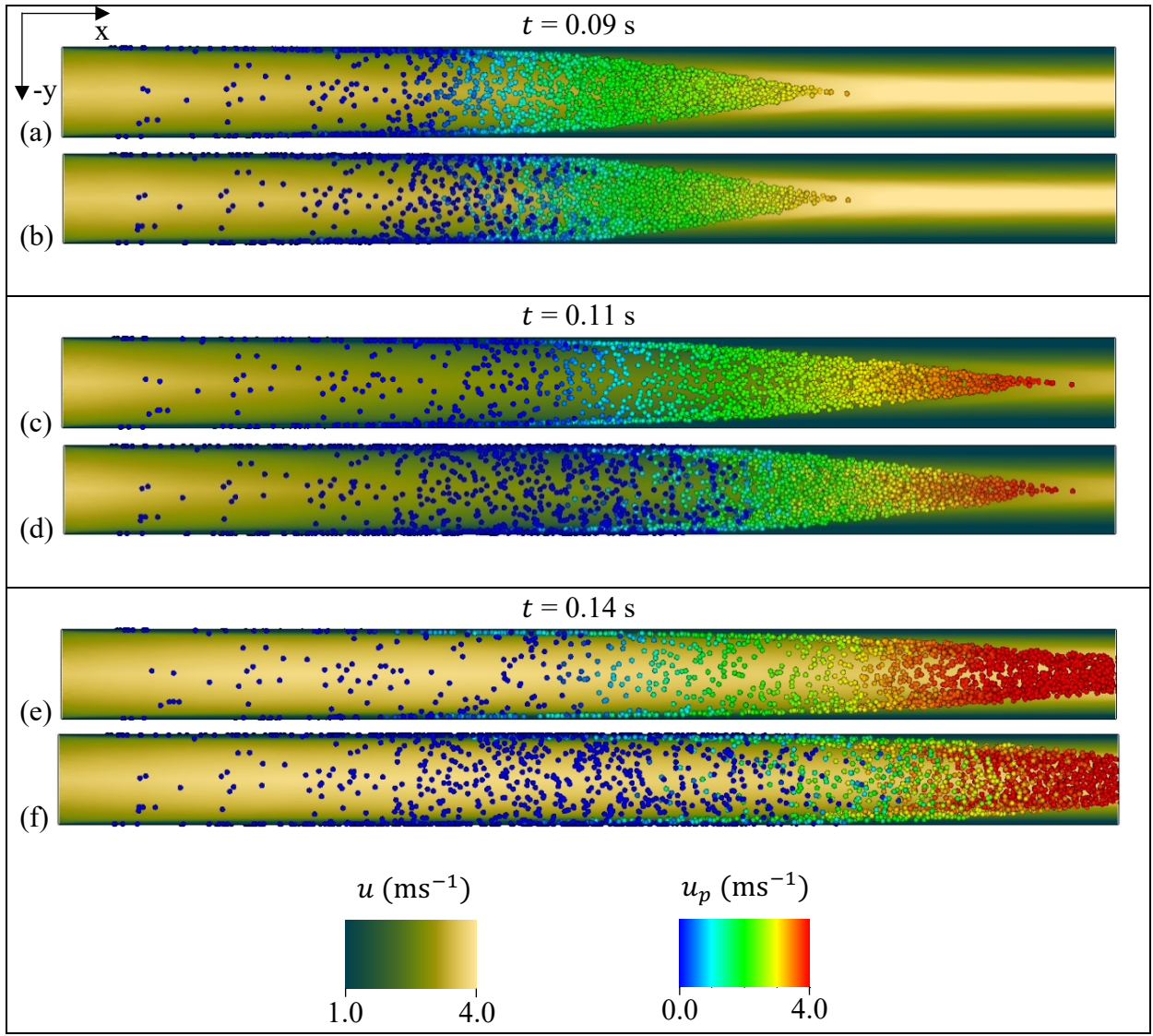

**Fig. 11.** Side view of simulation boundary for uncharged case and 5.04 fC case at different times. a, c, and e show the uncharged case at times t = 0.09 s, 0.11 s and 0.14 s respectively. b, d, and f show the 5.04 fC case at times t = 0.09 s, 0.11 s and 0.14 s respectively.

**Fig. 12** also confirms the hypothesis of an electrostatic force gradient. The gradient can be observed in the side view of the 5040 fC simulation (**Fig. 12(a)**). The tail of the plume (**Fig. 12(b)**) is populated by particles with a net negative electrostatic force (less than -1000 ms$^{-2}$) while the head of the plume (**Fig. 12(c)**) is populated by particles with a net positive electrostatic force (greater than 1000 ms$^{-2}$).

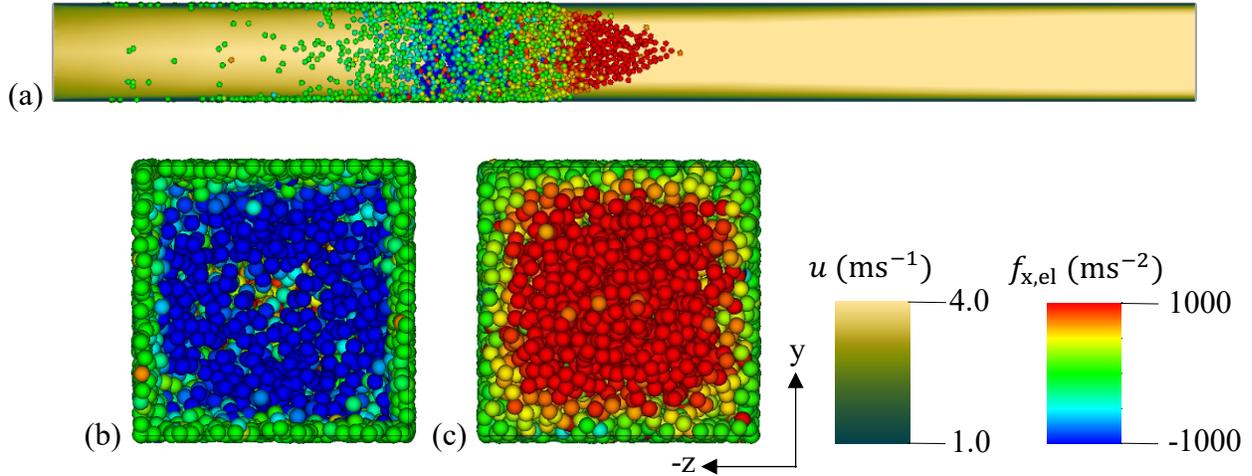

**Fig. 12.** Side view (a), inlet view (b), and outlet view (c) of 5040 fC simulation at $t = 0.062$ s. The inlet view shows particles at the tail end with negative streamwise electrostatic forces while the outlet view shows the particles at the head of the plume with positive electrostatic forces.

With the confirmation of the presence of electrostatic force gradient, the 5.04 fC simulation was checked for negative streamwise velocities. The presence of negative velocities would signify that the negative electrostatic forces were sufficient enough to cause backflow of particles. The absence of negative velocities in the 5.04 fC simulation was expected to signify absence of negative velocities in simulations with lower charge levels.

The probability distribution of streamwise particle velocity at different times for the 5.04 fC case was plotted in **Fig. 13**. The low-velocity tail of the distribution shifted, with time, to the left. Finally, at $t = 0.15$ s, a small fraction of the particles had negative velocities. This confirmed that the electrostatic forces were strong enough to cause backflow.

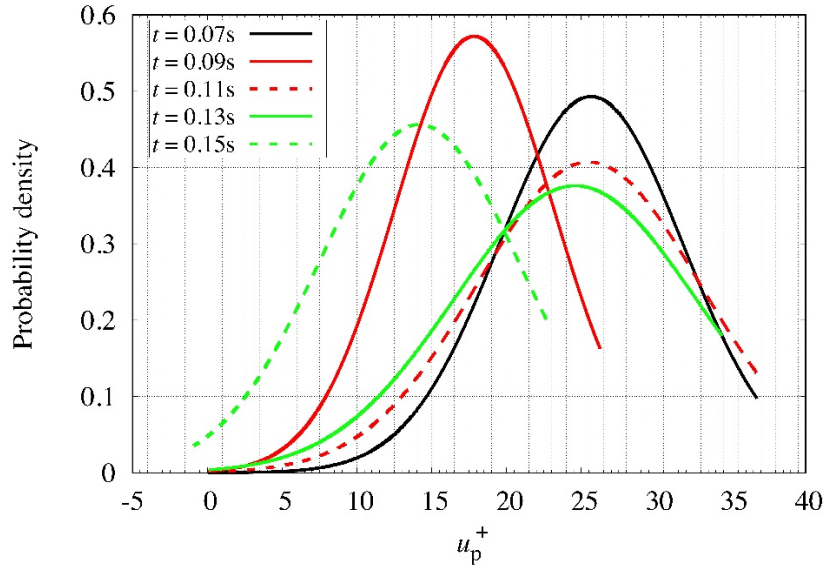

**Fig. 13.** Probability distribution of velocities at times t = 0.07 s, 0.09 s, 0.11 s, 0.13 s and 0.15 s, for the simulation of 5.04 fC per particle. A slightly negative value is observed at t = 0.15 s suggesting that particle backflow might be evident.

Following the discovery of negative velocity values, higher charge cases were simulated and analyzed. In **Fig. 14**, the velocity distributions for charge cases higher than 5.04 fC are shown at time $t = 0.15$ s. The extent of intrusion by the velocity distribution into the negative plane increased with increase in particle charge. This observation confirms that the flow reversal observed in pneumatic conveying systems can be linked to electrostatic forces induced by an electrostatic field in the system with, the reversal velocity positively correlating to the magnitude of the charge per particle. A snapshot of the particles in the 252 fC simulation confirming the presence of negative velocities is seen in **Fig. 15**.

**Fig. 16** analyzes the evolution of velocity distribution with time of the 5040 fC case. This evolution suggests that the presence of electrostatic force causes a wider and flatter spread of velocity values by increasing the maximum velocity and reducing the minimum velocity. As time progressed, the fastest particles left the duct, the maximum velocity in the distribution reduced and, the probability density of lower velocities increased.

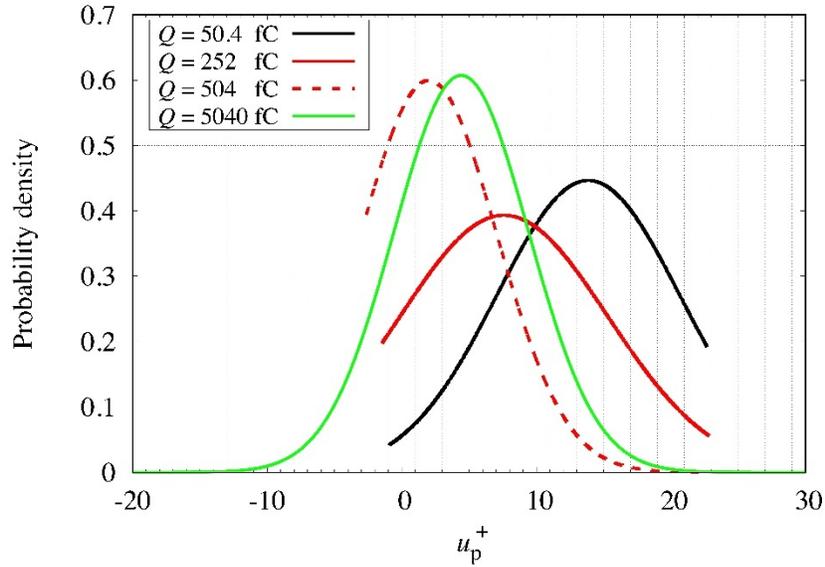

**Fig. 14.** Probability distribution of velocities at t = 0.15 s for 50.4 fC, 252 fC, 504 fC, and 5040 fC per particle. The minimum velocity decreases with charge confirming that particle backflow is indeed caused by the presence of electrostatic force and the reversal speed is positively correlated to the magnitude of the charge per particle.

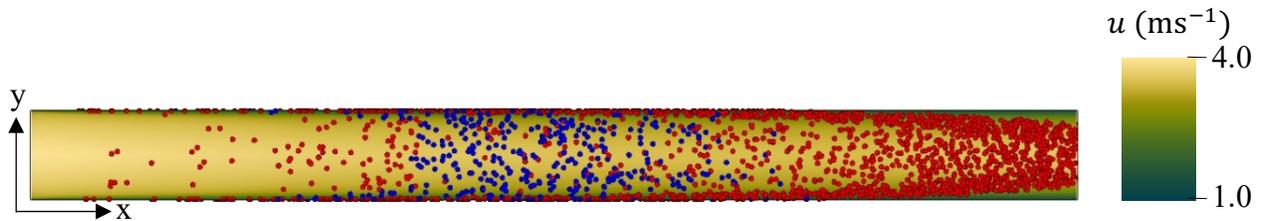

**Fig. 15.** Side view of 252 fC simulation at $t = 0.15$ s. The particles are colored based on the polarity of their velocity. Red particles having positive velocities and blue particles having negative velocities.

In conclusion, particle backflow can be observed in pneumatic conveying systems with discontinuous particle seeding (i.e., pulse injection) in the presence of an electrostatic field. The backflow velocity is dependent on the magnitude of the electrostatic force, and the charge on the particles.

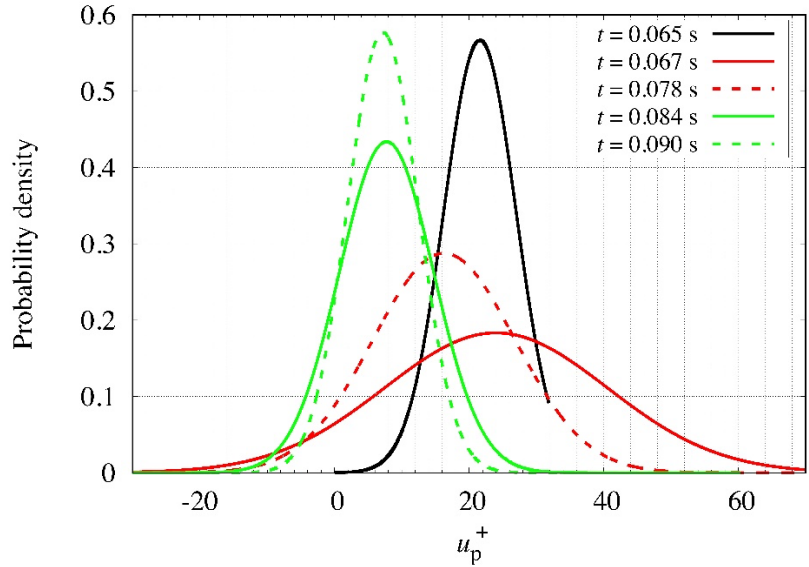

**Fig. 16.** Velocity distributions at times t = 0.065 s, 0.067 s, 0.078 s, 0.084 s and 0.090 s of simulation of 5040 fC per particle. Evolution shows initial widening and eventual leftward shift of distribution with progression of time. Widening caused by electrostatic repulsion and leftward shift caused by exit of faster particles from the outlet.

## 4. Conclusions and recommendations for future work

The primary objective of recreating particle backflow was achieved in the discontinuous seeding duct flow system for $St_{el}$ values greater than $8 \times 10^{-3}$. The magnitude of the backflow velocity increased with increase in the $St_{el}$ value. No backflow was observed in the continuous seeding duct flow and the channel flow between plates for up to an $St_{el}$ value of $8 \times 10^{-3}$

The presence of particle charge in channel flow and continuous seeding duct flow caused a migration of particles toward the walls of the boundary. The concentration of particles at the wall increased with the magnitude of the particle charge for both conveying systems. There was a decrease in the volume-average streamwise velocity with increase in particle for the channel flow but not for the continuous duct flow.

Now that electrostatics has been confirmed to be a factor for particle backflow, further work can be done to study particle backflow in polydisperse (both size and charge) particle systems and systems with in-situ particle charging.

## 5. Acknowledgements

Funding: This work was supported by the National Sciences and Engineering Research Council of Canada (NSERC), the PTB Guest Researcher Fellowship Grant and the European Research Council (ERC) under the European Union's Horizon 2020 research and innovation program (grant agreement No. 947606 PowFEct).

## Nomenclature

| | |
|---|---|
| $\boldsymbol{u}$ | fluid velocity, ms$^{-1}$ |
| $\rho$ | Fluid density, kgm$^{-3}$ |
| $p$ | Fluid dynamic pressure, Pa |
| $\nu$ | Fluid kinematic viscosity, m$^2$s$^{-1}$ |
| $F_\text{s}$ | Source term for momentum transfer from particle to fluid, m$^4$s$^{-2}$ |
| $F_\text{f}$ | Sink term for momentum loss due to friction at walls, m$^4$s$^{-2}$ |
| $\boldsymbol{u}_\text{p}$ | Particle velocity, ms$^{-1}$ |
| $\boldsymbol{u}_p^+$ | Normalized particle velocity |
| $\boldsymbol{f}_\text{fl}$ | Specific aerodynamic force, on particle, exerted by surrounding fluid, ms$^{-2}$ |
| $\boldsymbol{f}_\text{coll}$ | Specific collision force on particle, ms$^{-2}$ |
| $\boldsymbol{f}_\text{el}$ | Specific electrostatic force on particle, ms$^{-2}$ |
| $Q$ | Particle charge, C |
| $\boldsymbol{E}$ | Electric field strength around particle, NC$^{-1}$ |
| $\varphi_\text{el}$ | Electric potential, V |
| $\rho_\text{el}$ | Electric charge density, Cm$^{-1}$ |
| $\varepsilon$ | Electric permittivity, Fm$^{-1}$ |
| $\dot{N}_\text{in}$ | Number of particles fed per second, s$^{-1}$ |
| $t_\text{ref}$ | Reference time calculated for pulse seeding, s |
| $t_1$ | Duration of particle seeding pulse, s |
| $t_2$ | Time between two consecutive seeding pulses, s |
| $u_{bulk,t}$ | Fluid inlet centerline velocity, ms$^{-1}$ |
| $y_{\text{dim}}$ | Duct width, m |
| $\tau_1$ | Injection pulse factor |
| $\tau_2$ | Ratio of $t_2$ to $t_1$ |
| $St_{\text{el}}$ | Electrostatic Stoke's number |
| $u_\tau$ | Particle friction velocity, ms$^{-1}$ |
| $y^+$ | Dimensionless wall unit |
| $C$ | Normalized particle density |